\documentclass[epj]{svjour}
\usepackage{latexsym}
\usepackage{graphics}
\usepackage[usenames,dvipsnames]{xcolor} 

\begin{document}
\title{Kerr-Newman black holes can be generically overspun}
\author{Koray D\"{u}zta\c{s}\inst{1,2} 
}                     
%

\institute{Physics Department, Eastern Mediterranean  University, Famagusta, North Cyprus via Mersin 10, Turkey \and  Department of Natural and Mathematical Sciences,
\"{O}zye\u{g}in University, 34794 \.{I}stanbul Turkey}

\date{Received: date / Revised version: date}
%
\abstract{We construct thought experiments involving the perturbations of Kerr-Newman black holes by neutral test fields to evaluate the validity of the weak form of the cosmic censorship conjecture. We first show that neglecting backreaction effects, extremal Kerr-Newman black holes which satisfy the condition $(J^2/M^4)<(1/3)$ can be overspun by scalar fields. This result, which could not be discerned in the previous analyses to first order, is prone to be fixed by employing backreaction effects. However the perturbation of Kerr-Newman black holes by neutrino fields leads to a generic overspinning of the black hole due to the absence of a lower limit for the frequency of the incident wave to ensure that it is absorbed by the black hole. For this case, the destruction of the event horizon cannot be fixed by any form of backreaction effects. This result should not be interpreted as a counter-example to any of the previous results which were based on the assumption that the null energy condition is satisfied.
We clarify and justify our arguments with numerical examples.
%
} 
\maketitle
\section{Introduction}
The deterministic nature of general relativity was hampered by the development of singularity theorems by Penrose and Hawking. According to these theorems a singularity ensues as a result of gravitational collapse, given very reasonable assumptions~\cite{singtheo}. However if the gravitational collapse occurs in the way prescribed by Hawking and Penrose, the singularity is hidden behind an event horizon at the final state. In that case, the causal contact of the singularity with distant observers is disabled. Whether this can be generalised to include every type of gravitational collapse is an open problem. Penrose proposed the cosmic censorship conjecture to circumvent this problem. In its weak form (wCCC), the conjecture asserts that the gravitational collapse always ends up in a black hole surrounded by an event horizon~\cite{ccc}. Naked singularities must be forbidden in a physical universe. This way, the smooth structure of the space-time is maintained at least in the region outside the event horizon. The observers at the asymptotically flat infinity are not in causal contact with the singularity.

For decades, a concrete proof of the cosmic censorship conjecture has been elusive. In the absence of a concrete proof, Wald developed an alternative procedure to test the validity of the conjecture. In Wald type problems one starts with an extremal or a nearly extremal black hole with an event horizon surrounding the singularity. Then this black hole is perturbed by test particles or fields, which do not change the structure of the spacetime but lead to perturbations in mass, angular momentum, and charge parameters of the black hole. At the final stage one checks if it is possible to drive the black hole beyond extremality by the interaction with test particles or fields. In the first of these experiments Wald showed  that particles which carry sufficient charge or angular momentum to overcharge or overspin an extremal  Kerr-Newman black hole are not absorbed by the black hole~\cite{wald74}. The first thought experiment starting with a nearly extremal black hole instead of an extremal one was constructed by Hubeny~\cite{hu}. There, it was shown that a nearly extremal Reissner-Nordstr\"{o}m black hole 
can be overcharged by test particles. The same approach was adapted to overspin nearly extremal Kerr black holes by test particles\cite{Jacobson-Sot}. Many similar tests of wCCC were applied to  the black holes in Einstein-Maxwell theory ~\cite{f1,saa,gao,siahaan,magne,dilat,higher,v1}. For some cases, it was shown that the destruction of the horizon can be prevented by employing backreaction effects~\cite{back1,hu1,back2,w2}. The possibility to destroy the horizon by quantum tunnelling of particles was analysed~\cite{q1,q2,q3,q4,q5,q6,q7}. For the asymptotically anti de-Sitter case, Rocha and Cardoso concluded that it is not possible to overspin a Banados, Teitelboim, Zanelli (BTZ) black hole after an analysis which is restricted to the case of extremal black holes~\cite{vitor}. Their conclusion was supported in the following works~\cite{zhang,rocha,gwak1,gwak2}. However we have shown that overspinning is possible by using test particles and fields test if we start with a nearly extremal black hole instead~\cite{btz}. 

Another intriguing problem is to test the validity of wCCC in the case of test fields scattering off black holes. After the pioneering work of Semiz on the possibility of destroying a dyonic Kerr-Newman black hole by the interaction with scalar fields~\cite{semiz}, many thought experiments involving the perturbations of space-times by test fields were constructed~\cite{toth1,emccc,overspin,sh,duztas,toth,natario,duztas2,mode,thermo,taub-nut,gwak3,kerrsen}. In this work we perturb Kerr-Newman black holes by  neutral test fields and investigate if they can be overspun into naked singularities. We first evaluate the case of scalar fields interacting with extremal Kerr-Newman black holes.  We perform a second order analysis to show that the extremal black holes with sufficiently low angular momentum (to be made precise) can indeed be overspun by scalar fields into naked singularities.  It is not --therefore has not been-- possible to notice the overspinning of extremal black holes in  similar analyses of the first order~\cite{w2,natario}. However, the destruction of extremal black holes by scalar fields can be interpreted as an intermediate result which is likely to be fixed by employing backreaction effects in a full second order analysis. 

In the scattering of bosonic fields, there exists a lower limit for the frequency of the incoming  field to allow its absorption. However, such a limit does not exist for fermionic fields which allows the absorption of mode with low energy and relatively high angular momentum. In section (\ref{neutrino}) we evaluate the interaction of Kerr-Newman black holes with neutrino fields which leads to a generic destruction of the event horizon. We clarify and justify our arguments with numerical examples.  
\section{Scalar fields, Kerr-Newman black holes and wCCC}
The Kerr-Newman metric describes a black hole with an event horizon surrounding the singularity if the mass $(M)$, angular momentum ($J=Ma$) and charge $(Q)$ parameters of the spacetime satisfy the inequality
\begin{equation}
M^2-Q^2-a^2 \geq 0 \label{crit}
\end{equation}
In Wald type problems one starts with a black hole satisfying (\ref{crit}). Then the spacetime parameters are perturbed to check if the black hole can be driven beyond extremality. In the problems involving particles one first demands that the test particle crosses the horizon to be absorbed by the black hole. This condition gives us the minimum value for the energy of the test particle, which contributes to the mass parameter of the spacetime. The maximum value for the energy is derived by demanding that (\ref{crit}) is violated at the end of the interaction so that the spacetime parameters represent a naked singularity. 

In the problems involving fields we envisage a test field that is incident on the black hole from infinity. After the interaction the field decays away and the spacetime parameters attain their final values. For bosonic fields, there exists a lower bound for the frequency of the incoming field analogous to the minimum energy of the particle, which is the limiting frequency for superradiance to occur. If the frequency of the field is lower than the superradiance limit, the field will not be absorbed by the black hole. It will scatter back to infinity with a larger amplitude borrowing the access energy from the angular momentum of the black hole. In this case (\ref{crit}) will be reinforced rather than challenged. Similar to the particle case we derive the maximum frequency for the field by demanding that (\ref{crit}) is violated at the end of the interaction. 
 
In this section we perturb  extremal and nearly extremal Kerr-Newman black holes with neutral scalar fields that have frequency $\omega$ and azimuthal wave number $m$. At the end of the interaction the mass and angular momentum parameters of the black hole are modified.
\begin{eqnarray}
&&M_{\rm{fin}}=M +\delta M \nonumber \\
&& J_{\rm{fin}}=J +\delta J=J+(m/\omega) \delta M  \nonumber \\ 
&&Q_{\rm{fin}}=Q
\end{eqnarray}
where $\delta M=\delta E$ is the energy of the incoming field, and $\delta J$ is its angular momentum. The charge of the black hole is invariant since we work with neutral fields. We investigate if it is possible to find real values for  the frequency of the incoming scalar field so that these two conditions are simultaneously satisfied: (i) the field is absorbed by the black hole (ii) the final parameters of the spacetime represent a naked singularity violating the inequality (\ref{crit}).
\subsection{Overspinning extremal  black holes}
By definition, an extremal Kerr-Newman black hole satisfies
\begin{equation}
\delta_{\rm{in}}\equiv M^2-Q^2-\frac{J^2}{M^2}=0 \label{param}
\end{equation}
where we have defined $\delta_{\rm{in}}$. We perturb this black hole with a scalar field. We demand that (\ref{crit}) is violated at the end of the interaction, i.e. $\delta_{\rm{fin}}<0$
\begin{equation}
\delta_{\rm{fin}}\equiv (M+\delta M)^2-Q^2-\frac{(J+\delta J)^2}{(M+\delta M)^2}<0 \label{overspin1}
\end{equation}
We choose $\delta E=\delta M=M\epsilon$ for the scalar field, where $\epsilon \ll 1$. We eliminate $(M^2-Q^2)$ from (\ref{overspin1}) using (\ref{param}). (\ref{overspin1}) takes the form
\begin{equation}
M^2 \left( \epsilon^2 + 2 \epsilon + \frac{J^2}{M^4}\right) < \frac{\left( J+ \frac{m}{\omega}M \epsilon \right)^2}{M^2(1+\epsilon)^2}
\end{equation}
We take the square root of both sides and define the dimensionless variable $\alpha\equiv (J/M^2)$. Elementary algebra yields that the condition $\delta_{\rm{fin}}<0$ is equivalent to
\begin{equation}
\omega < \omega_{\rm{max-ex}}=\frac{m\epsilon}{M\left[ (1+\epsilon)\sqrt{\epsilon^2 + 2\epsilon + \alpha^2} - \alpha \right] } \label{wmaxex}
\end{equation}
To ensure that $\delta_{\rm{fin}}$ is negative at the end of the interaction, i.e. the extremal Kerr-Newman black hole is overspun into a naked singularity, the frequency should be below $\omega_{\rm{max-ex}}$ given in (\ref{wmaxex}) if the energy of the incoming field is chosen to be $\delta E=M\epsilon$. However these conditions are not sufficient. The frequency should also be above the superradiance limit so that the scalar field is absorbed by the black hole. Since $r_+=M$ for an extremal black hole, the limiting frequency for superradiance for neutral fields is given by
\begin{equation}
\omega_{\rm{sl-ex}}=\frac{ma}{r_+^2 + a^2}=\frac{m}{M\left( \frac{1}{\alpha}+ \alpha\right)} \label{wslex}
\end{equation}
The frequency of the incoming field should be larger than the superradiance limit to ensure its absorption. 
If the limiting frequency for superradiance is lower than the maximum frequency derived in (\ref{wmaxex}) for any values of $\alpha$ and $\epsilon$, it can be possible to overspin an extremal Kerr-Newman black hole into a naked singularity by using test scalar fields. The condition $\omega_{\rm{sl-ex}}<\omega_{\rm{max-ex}}$ is equivalent to
\begin{equation}
\epsilon\left( \frac{1}{\alpha} + \alpha \right) + \alpha> (1+\epsilon)\sqrt{\epsilon^2 + 2\epsilon +\alpha^2}
\end{equation}
We take the square both sides and keep the terms up to second order in $\epsilon$. This leads to 
\begin{equation}
\alpha^2<\frac{1}{3} \label{condiex}
\end{equation}
If the condition (\ref{condiex}) is satisfied by an extremal Kerr-Newman black hole, the limiting frequency for superradiance is less than the maximum frequency that can be chosen to overspin the black hole. Then, if we choose $\delta E=M\epsilon$ and   $\omega_{\rm{sl-ex}}<\omega<\omega_{\rm{max-ex}}$ for the incoming field, the extremal Kerr-Newman black hole will be overspun into a naked singularity at the end of the interaction. 
\subsection{Comparison with previous results}
 Recently Natario, Queimada, and Vicente (NQV)   \cite{natario}, and Sorce and Wald (SW) \cite{w2} claimed that test fields satisfying the null energy condition cannot destroy extremal Kerr-Newman black holes. In both works the authors derive the condition which is required to ensure that a test particle or field is absorbed by the black hole.
\begin{equation}
\delta M -\Omega \delta J - \Phi \delta Q \geq 0 \label{needham}
\end{equation}
This condition was first derived by Needham without assuming cosmic censorship~\cite{needham}. Needham's condition (\ref{needham}) reduces to $\omega \geq \omega_{\rm_{sl}}$ for test fields with $\delta J=(m/\omega) \delta M$, i.e. there is no 
contradiction with the condition imposed in this work. We agree that the frequency of the test field should be larger than the superradiance limit if it is absorbed by the black hole. SW argue that a violation of cosmic censorship will occur if the perturbation of the extremal black hole satisfies
\begin{equation}
2M\delta M < 2(J/M)(M\delta J -J\delta M)/M^2 +2Q\delta Q \label{condisr}
\end{equation}
However in Eq. (\ref{overspin1}), for overpinning to occur we demand that
\begin{equation}
M_{\rm{fin}}^2-Q_{\rm{fin}}^2-(J_{\rm{fin}}^2)/(M_{\rm{fin}}^2)<0 \label{condiduz}
\end{equation}
which  accurately determines the condition that the final state represents a naked singularity. Eventually, we have to calculate $M_{\rm{fin}}^2$ and $J_{\rm{fin}}^2$ to decide whether or not the event horizon is destroyed. For that reason even though we ignored the backreaction effects in this work, we did not neglect the terms $(\delta M)^2$ and $(\delta J)^2$. That is the main difference between this work and the previous works by NQV and SW, regarding the perturbations of extremal black holes satisfying the null energy condition.

The main result of SW is that the conditions (\ref{needham}) and (\ref{condisr}) cannot be satisfied simultaneously. Based on this result, they conclude that extremal black holes cannot be destroyed. However to decide whether or not the horizon is destroyed, one should check the validity of the condition (\ref{condiduz}), rather than  (\ref{condisr}).  Therefore, unless one ignores the second order terms $(\delta M)^2$, $(\delta J)^2$, one cannot infer that the event horizon will be preserved from the result of SW. NQV also apply a first order analysis to conclude that the final mass is above the mass of  a corresponding extremal black hole so that extremal black holes cannot be destroyed. The validity of their results is also restricted to the case $(\delta M)^2 \to 0$, $(\delta J)^2 \to 0$, and $(\delta Q)^2 \to 0$.

Let us clarify our arguments with a numerical example. Consider an extremal black hole with $M=1$, and $\alpha=0.3$. The extremality condition yields that $Q^2=0.91$. For this black hole $\alpha^2 <(1/3)$, so according to the analysis in this section, the limiting frequency for superradiance should be less than the maximum frequency derived in (\ref{wmaxex}). Choosing $\epsilon=0.01$ one numerically verifies that this is indeed the case.
\begin{eqnarray}
&& \omega_{\rm{sl}}=0.275229(m/M) \nonumber \\
&& \omega_{\rm{max-ex}}=0.284646(m/M)
\end{eqnarray}
Let us perturb this black hole with a neutral scalar field with:
\begin{eqnarray}
&& \omega=0.276 (m/M) \nonumber \\
&& \delta M=M\epsilon=0.01 M \nonumber \\
&& \delta J= (m/\omega) \delta M=0.036232 M^2 \nonumber \\
&&\delta Q=0
\end{eqnarray}
This perturbation satisfies the Needham's condition ($\omega>\omega_{\rm{sl}}$), which means that it will be absorbed by the black hole. In accord with the result of SW it does not satisfy (\ref{condisr}), since 
\begin{eqnarray}
& &2M\delta M=0.02  \nonumber \\
& &[2(J/M)(M\delta J -J\delta M)]/(M^2)=0.019939 \nonumber \\
&\Rightarrow & 2M\delta M >[2(J/M)(M\delta J -J\delta M)]/(M^2)
\end{eqnarray}

At this stage one would conclude that wCCC cannot be violated in a first order analysis. However, the precise  calculation of $\delta_{\rm{fin}}$ with the same perturbation, yields that
\begin{eqnarray}
\delta_{\rm{fin}}&=& M_{\rm{fin}}^2-Q_{\rm{fin}}^2-(J_{\rm{fin}}^2)/(M_{\rm{fin}}^2) \nonumber \\
&=&(1+0.01)^2 -0.91-\frac{(0.3 + 0.036232)^2}{(1+0.01)^2} \nonumber \\
&=&-0.000724 \label{deltafinex}
\end{eqnarray}
The negative sign in (\ref{deltafinex}) indicates that the extremal black hole is overspun into a naked singularity. In other words, neglecting backreaction effects, extremal Kerr-Newman black holes for which $\alpha^2<(1/3)$, can be overspun into naked singularities by neutral scalar fields with a judicious choice of frequency. This result cannot be discerned in a first order analysis. However the magnitude of $\delta_{\rm{fin}}$ in (\ref{deltafinex}) suggests that it can be fixed by  employing backreaction effects. In fact, for nearly extremal Kerr-Newman black holes SW derived an inequality for second order variations As we have mentioned in the introduction, the over-spinning of extremal Kerr-Newman black holes by scalar fields is merely an intermediate result the validity of which is limited to the case where one ignores backreaction effects.  One can also calculate the second order variations  for extremal Kerr-Newman black holes or employ an alternative method to incorporate the backreaction effects, to restore the event horizons of extremal Kerr-Newman black holes.

\section{Neutrino fields and wCCC}\label{neutrino}
It is known that the superradiance does not occur for neutrino fields. (see e.g. \cite{chandra}). For that reason the lower limit for the frequency  to ensure the absorption of the test field does not exist. The energy-momentum tensor for neutrino fields does not satisfy the null energy condition, either. Therefore, neutrino fields do not obey the Needham's condition (\ref{needham}) to be absorbed by a black hole. Either by the argument of the absence of superradiance or by the violation of Needham's condition, every mode will be absorbed when neutrino fields scatter off Kerr-Newman black holes. This leads to drastic results as far as cosmic censorship is concerned. If the absorption of modes with lower frequency is allowed, their contribution to the angular momentum will be much larger than the contribution to the mass parameter of the black hole. In that case overspinning becomes robust and it also applies to extremal Kerr-Newman black holes which does not satisfy (\ref{condiex}). 

The scattering of neutrino fields should not be confused with the thought experiments involving the tunnelling of a single fermion~\cite{q3,q4,q5,q6}. The evaporation of black holes dominates the effect of a single particle by many orders of magnitude~\cite{q6,q7}, whereas its effect is negligible against challenging fields~\cite{duztas2}. We argued this in detail in \cite{mode}. (See section IV in \cite{mode})

For a numerical example, let us consider an extremal black hole with $M=1$ and $\alpha=0.6$. This black hole does not satisfy the condition (\ref{condiex}), therefore it cannot be destroyed by scalar fields even if one ignores the backreaction effects. In particular with $\epsilon=0.01$, one derives that
\begin{eqnarray}
&&\omega_{\rm{sl}}=0.441176(m/M) \nonumber \\
&&\omega_{\rm{max}}=0.440767(m/M)
\end{eqnarray}
Since $\omega_{\rm{sl}}>\omega_{\rm{max}}$, one cannot find a frequency for the incoming wave that will be absorbed by the black hole to overspin it, provided that the perturbation satisfies the null energy condition or it is subject to superradiance. However, the lower bound for frequency does not exist for neutrino fields. One can choose any frequency below $\omega_{\rm{max}}$ to overspin the black hole. Consider a neutrino field with frequency $\omega=0.2(m/M)$ and energy $\delta M=0.01M$, i.e. $\epsilon=0.01$.
\begin{eqnarray}
&&\omega=0.2(m/M) \nonumber \\
&&\delta M=0.01M  \nonumber \\
&& \delta J=(m/\omega) \delta M=0.05 M^2 \nonumber \\
&& \delta Q=0 \label{neutfield}
\end{eqnarray}
This field will be absorbed by the black hole, since superradiance does not occur, or the Needham's condition (\ref{needham}) does not apply. For such low energy modes, the relative contribution to angular momentum is enhanced as $\omega$ is lowered. One can calculate $\delta_{\rm{fin}}$ 
\begin{eqnarray}
\delta_{\rm{fin}}&=& (1+0.01)^2 - 0.64 - \frac{(0.6 + 0.05)^2}{(1+0.01^2} \nonumber \\
&=&-0.034075 \label{deltafinneutex}
\end{eqnarray}
The value of $\delta_{\rm{fin}}$ represents a generic violation of wCCC. We observe that $\vert \delta_{\rm{fin}} \vert \gg M^2\epsilon^2 $. This robust violation cannot be fixed by any form of backreaction effects. 
\subsection{Neutrino fields and nearly extremal black holes}
As we mentioned above, SW derived an inequality for the second order variations of nearly extremal Kerr-Newman black holes and concluded that the event horizon cannot be destroyed if one employs backreaction effects. However these arguments do not apply to neutrino fields since their energy momentum tensor does not satisfy  the null energy condition. In this section we attempt to overspin nearly extremal Kerr-Newman black holes by neutrino fields. The main difference from bosonic fields is the absence of a lower limit to ensure the absorption of the incoming field. Let us consider a nearly extremal Kerr-Newman black hole parametrized as
\begin{equation}
\delta_{\rm{in}}\equiv M^2-Q^2-\frac{J^2}{M^2}=M^2 \epsilon^2 \label{param1}
\end{equation}
where $\epsilon \ll 1$. We perturb this black hole with a neutrino field. We demand that (\ref{crit}) is violated at the end of the interaction, i.e. $\delta_{\rm{fin}}<0$. Again we choose $\delta E=\delta M=M\epsilon$ for the incoming field.  Using (\ref{param1}), the condition $\delta_{\rm{fin}}<0$ can be expressed in the form
\begin{equation}
M^2 \left( 2\epsilon^2 + 2 \epsilon + \frac{J^2}{M^4}\right) < \frac{\left( J+ \frac{m}{\omega}M \epsilon \right)^2}{M^2(1+\epsilon)^2}
\end{equation}
Proceeding in the same way  as the extremal case, we take the square root of both sides and define the dimensionless variable $\alpha\equiv (J/M^2)$. Elementary algebra yields that the condition $\delta_{\rm{fin}}<0$ is equivalent to
\begin{equation}
\omega < \omega_{\rm{max}}=\frac{m\epsilon}{M\left[ (1+\epsilon)\sqrt{2\epsilon^2 + 2\epsilon + \alpha^2} - \alpha \right] } \label{wmaxnex}
\end{equation}
The lower bound for frequencies does not exist for neutrino fields, therefore any field with $\omega<\omega_{\rm{max}}$ will be absorbed by the nearly extremal Kerr-Newman black hole to overspin it into a naked singularity. If the frequency of the incoming field is slightly lower than $\omega_{\rm{max}}$, the absolute value of $\delta_{\rm{fin}}$ will be of the order $M^2\epsilon^2$, and the second order variations studied by SW will be able to restore the horizon. However, as the frequency is lowered further, the absolute value of $\delta_{\rm{fin}}$ will be of the order $M^2\epsilon$ leading to a generic overspinning which cannot be fixed by any form of backreaction effects.

To clarify the arguments above let us start with a nearly extremal Kerr-Newman black hole with $M=1$ and $\alpha=0.6$. Choosing $\epsilon=0.01$, the parametrization  (\ref{param1}) implies that $Q^2=0.6399$. To overspin this black hole the frequency of the incoming field has to be below the maximum value given in (\ref{wmaxnex}).
\begin{equation}
\omega<\omega_{\rm{max}}=0.439181(m/M)
\end{equation}
There is no lower limit for $\omega$, as far as neutrino fields are concerned. Let us perturb the nearly extremal Kerr-Newman black hole with the same neutrino field described in (\ref{neutfield}).  One can calculate $\delta_{\rm{fin}}$
\begin{eqnarray}
\delta_{\rm{fin}}&=& (1+0.01)^2 - 0.6399 - \frac{(0.6 + 0.05)^2}{(1+0.01^2} \nonumber \\
&=&-0.033975 \label{deltafinneut}
\end{eqnarray}
The value of  $\delta_{\rm{fin}}$ represents a generic overspinning of the nearly-extremal Kerr-Newman black hole, which cannot be fixed by considering the second order variations studied by SW or employing any form of backreaction effects. In particular, the contribution of the backreaction effects will be of the order $M^2 \epsilon^2$, whereas $\vert \delta_{\rm{fin}} \vert \sim M^2 \epsilon$. The nearly extremal Kerr-Newman black holes can also be generically overspun into naked singularities by neutrino fields. 
\section{Conclusions}
In this work  we constructed thought experiments in which neutral test fields scatter off Kerr-Newman black holes to check whether the black holes can be overspun into naked singularities. We first investigated the possibility of overspinning Kerr-Newman black holes by scalar fields. Though we neglected the backreaction effects, we retained the second order terms $(\delta J)^2$ and $(\delta M)^2$. Therefore our results differ from the previous analyses to first order by SW and NQV. We  showed that --neglecting backreaction effects-- extremal Kerr-Newman black holes which satisfy $\alpha^2=(J^2/M^4)<(1/3)$ can be overspun by scalar fields. However, our numerical calculation suggests that the overspinning of extremal black holes by scalar fields is likely to be fixed by employing backreaction effects. In particular SW have already proved this for nearly extremal Kerr-Newman black holes by incorporating  the effect of the second order variations which account for the self-force effects.  

The generic overspinning of  Kerr-Newman black holes occurs in the interactions with neutrino fields. In this case, superradiance does not occur, and the energy-momentum tensor does not satisfy the null energy condition, which implies that the modes with very low energy and relatively high angular momentum can also be absorbed by the black hole. This leads to a generic destruction of the event horizon. We applied a numerical calculation to clarify this argument. The values of $\delta_{\rm{fin}}$ in (\ref{deltafinneutex}) and (\ref{deltafinneut}) indicate that the backreaction effects --which were neglected in this work-- cannot compensate for this generic overspinning. The contribution of the backreaction effects will be in second order, whereas the absolute value of $\delta_{\rm{fin}}$ is in the first order in the numerical examples (\ref{deltafinneutex}) and (\ref{deltafinneut}).
In fact, one can  lower the frequency of the incoming field and increase the absolute value of $\delta_{\rm{fin}}$ even further. However, the test field approximation can be distorted in the interaction with such a field which would considerably increase the angular momentum parameter of the background.

The generic destruction of the event horizon by neutrino fields derived in this work does not constitute a counter-example to any of the previous results which were based on the assumption that the null energy condition is satisfied. The absence of a lower limit for the absorption of neutrino fields leads to analogous results in the cases of Kerr~\cite{q7}, BTZ~\cite{btz}, Kerr-Taub-NUT~\cite{taub-nut}, and Kerr-Sen \cite{kerrsen} black holes. 

 We should also note that the treatment of neutrino fields in this work is purely classical. In the classical picture, the absorption probability is positive for all modes of neutrino fields. (See e.g.~\cite{chandra} and~\cite{page} for explicit calculation of these probabilities.) That is the main reason which leads to drastic results as far as cosmic censorship is concerned. A quantum analysis may well yield a different result to fix for the overspinning of the black holes. In the quantum picture one should also take into account the evaporation of black holes. Both in the form of Hawking radiation and the spontaneous emission previously studied by Zeldovich~\cite{zeldo}, Starobiinski~\cite{staro}, and Unruh~\cite{unruh}, the evaporation of black holes work in favour of cosmic censorship. As  extremality is approached the black holes emit particles in the modes $\omega<m\Omega$, which decreases the angular momentum of the black hole more than its mass, and carries the black hole away from extremality. Though, this evaporation dominates the effect of a single particle by many orders of magnitude~\cite{q6,q7}, its effect is negligible against challenging fields~\cite{duztas2,mode}. Therefore, it appears that we should execute a quantum analysis of fermionic scattering beyond the semi-classical level, to preserve the validity of cosmic censorship. 
%
%

\end{document}